\documentclass[a4paper]{jpconf}
\usepackage{graphicx}
\usepackage{color}
\begin{document}

\providecommand{\ttbar}{\rm t\bar{t}}
\providecommand{\ttbarH}{\rm t\bar{t}H}
\providecommand{\mtop}{m_{\rm t}}
\providecommand{\pT}{p_{\rm T}}
\providecommand{\kT}{k_{\rm T}}
\providecommand{\GeV}{~\rm GeV}
\providecommand{\TeV}{~\rm TeV}
\providecommand{\fbinv}{~\rm fb^{-1}}
\providecommand{\abinv}{~\rm ab^{-1}}
\providecommand{\wwbbbar}{~{\rm W}b{\rm W}\bar{b}}
\providecommand{\FIXME}[1]{{\color{red}#1}}

\title{Perspectives on top quark physics after Run I of the LHC: $\sqrt{s}=13\TeV$ and beyond}

\author{Pedro Ferreira da Silva\\
{\it on behalf of the ATLAS and CMS collaborations
\footnote{Presented at the 7$^{\rm th}$ International Workshop on Top Quark Physics,
29$^{\rm th}$ September to 3$^{\rm rd}$ October, 2014 - Cannes, France}
}\
}

\address{CERN, PH department, CH-1211 Geneva 23 Switzerland}
\ead{psilva@cern.ch}

\begin{abstract}
A summary of the on-going preparations from the ATLAS and CMS collaborations
to perform top quark physics in Run II of the LHC and at the HL-LHC is given.
To maintain the current level of precision and profit from the high-luminosity scenario
expected in the next runs of the LHC, several new reconstruction techniques and 
detector upgrades are foreseen. The prospects for precise measurements and 
possible discovery stories for new physics with top quarks are summarized.
\end{abstract}

\section{Introduction}
\label{sec:introduction}

The first run of proton-proton collisions at the CERN Large Hadron Collider (LHC) 
has brought the confirmation of the existence of a scalar boson which resembles,
so far, the one which breaks the electroweak symmetry (EW) 
in the standard model (SM) of particle physics.
Despite being an outstanding discovery, there are many shortcomings of the SM
which seek to be solved and for which the heaviest of the elementary particles,
the top quark (t), may be the key.

Given its high mass ($\mtop\approx173\GeV$), 
the top quark is expected to dominate the virtual corrections to the Higgs boson mass.
These corrections happen to balance against 
the ones received from the other known elementary particles with an unnatural precision
which is inversely proportional to the square of the energy scale up to which the SM is valid.
If such scale is the Planck scale, the cancellation would be $\sim 10^{-34}$.
In addition, the interplay between the recently discovered boson and the top quark
is expected to affect the structure of the EW vacuum.
Similarly, if the SM is extrapolated up to the Planck scale,
the Higgs potential is  expected to be critically affected by the value of $\mtop$.
Although possible, such strong relations between the SM and the Planck scales,
may hint that physics beyond the SM (BSM), 
could be contributing to the Higgs mass and self-interactions.
In several of the possible BSM extensions which may ease these strong dependencies,
the top quark couples preferentially to new partners or interaction mediators.
In addition to this privileged role in the EW symmetry breaking mechanism, and BSM scenarios,
the simple properties of the top quark, 
$\mtop$ and the Cabibbo-Kobayashi-Maskawa (CKM) matrix elements,
have direct consequences on the SM predictions for the W mass 
and the branching ratio of the golden $B_s\rightarrow \mu^+\mu^-$ decay channel,
respectively. Both cases have high sensitivity to BSM contributions.

From these examples it becomes clear that the top quark plays a unique role in the particle physics.
Despite the achieved precision in the SM measurements
and the limits in the direct searches, after the LHC Run I,
it hasn't been yet possible to learn which BSM physics is in reach of the top quark~\cite{Schwanenberger}.
Further progress can only be achieved by analyzing more data
\footnote{
Although the topic is beyond the scope of this writeup, one should notice
that the integrated luminosity acquired in Pb-Pb collisions, wasn't yet enough
to use top quarks to probe the quark-gluon plasma in Run I.
Nevertheless that is foreseen to be feasible in the upcoming heavy ion runs of the LHC
enlarging the scope of interest in top quark physics~\cite{cmsecfa}.
}.
Thus there are strong motivations to keep pursuing its physics in the next LHC runs,
both expanding measurements of its properties as well as searching for BSM physics produced in association with, or mimicked by, top quarks.
A review of the prospects for top quark physics in the next runs is given in the next sections of this run-through,
based on~\cite{snowmass,atlaspublic:twiki,cmspublic:twiki,ATLAS:2013hta,cmsecfa}.

\section{Challenges for $\sqrt{s}=13\TeV$ and beyond}
\label{sec:challenges}

\subsection{The machine and the detectors}
\label{subsec:machineanddetectors}

The upcoming Run II of the LHC will bring up to $300\fbinv$ of pp collisions
at $\sqrt{s}=13\TeV$. This will yield a drastic increase of the production cross section
of strongly produced processes, as the parton luminosity, in particular gluon-gluon, 
increases very steeply: $\mathcal{O}(10)$ gain is expected for $M_{\rm gg}>1\TeV$.
The increase in energy and integrated luminosity will be accompanied by an increase in pileup
(which may reach up to 50 interactions per bunch crossing),
and a reduced bunch spacing of 25 ns.
The ultimate phase of the LHC,
the High Luminosity LHC (HL-LHC), is targeted to start in 2024 and it is expected to
collect up to $3\abinv$ of pp collision data at 14 TeV, after 10 years of operation.
Targeting an instantaneous luminosity of $5\cdot 10^{34} {\rm cm^{-2} s^{-1}}$ the average pileup will be close to 140 
interactions per bunch crossing.
In order to be able to produce such a large dataset the LHC machine will have to undergo a series
of upgrades which will enable, amongst others, 
the replacement of the dipoles, the upgrade of collimation and cryogenics,
the installation of crab cavities and the leveling of luminosity.
To fully profit from this scenario both ATLAS~\cite{ATLAS} and CMS~\cite{CMS} detectors will undergo through a series
of upgrades which started already for Run II with a completion of the muon coverage
and muon trigger systems and installation of smaller radius beam pipes.
In addition CMS replaced the PMTs and SiPMs of the hadronic calorimeter and ATLAS installed
an additional pixel layer and a diamond beam monitor.
Towards the HL-LHC the forward calorimeters and the inner trackers will be replaced gradually
as well as the electronics of most sub-detectors and the trigger systems.

\subsection{Taming pileup}
\label{subsec:pileup}

To explore fully the physics in reach of the next runs of the LHC
detectors have to face inevitably the pileup challenge.
Amongst other effects, the incoherent overlap of energy flux from both in-time and out-of-time
bunches contaminates the isolation of leptons, the clustering of jets and the
measurement of the missing transverse energy in the event.
As an example, the jet energy resolution is expected to be drastically worsened,
degrading by a factor of 2 for $\pT \sim 30\GeV$ when the average pileup is 140.
Furthermore, non-negligible contamination of the events with
fake jets, from pileup, is expected:
even if pileup jets are characterized
by a steeply falling spectrum in $\pT$, 
their rate is proportional to the average number of simultaneous collisions occurring each bunch crossing.
If no new techniques are applied to mitigate these effects it will become impossible to 
achieve the physics goals aimed for Run II and the HL-LHC.

Using the association of tracks to the primary vertexes (PV) reconstructed in an event
has already proved to be a useful technique to mitigate in-time pileup during Run I.
Either by using the distribution of tracks associated to the leading-$\pT$ PV of the event
as a pileup jet discriminator (ATLAS' Jet Vertex Tagger), 
either by removing the tracks associated to the non-leading $\pT$ PV
before running the jet clustering algorithm (CMS's Charged Hadron Subtraction).
After correcting for inefficiencies induced by hit confusion, fake hits,
vertex merging, etc., stable identification efficiencies and improvement 
in the jet energy resolution are obtained.
At calorimeter level, pileup can be further tamed by
exploring the granularity in the calorimeters.
By relaxing iteratively the requirement on the hit significance above noise,
when building energy clusters, the Topological Clustering algorithm used in ATLAS,
proves to be efficient in reducing both noise and pileup, prior to jet reconstruction.
Pileup subtraction can also be carried by using constituent subtraction with the jet area, 
which can be enriched by using tracking information.
In these techniques it is important to tune the thresholds and weights assigned to each sub-detector 
in the clustering algorithms. These may depend on the pileup scenario and also require a good simulation of noise, 
calorimeter aging, zero suppression and selective readout algorithms.

An interesting technique arises from the usage of $\kT$-like distances to quantify
the probability of nearby energy deposits, tracks or fully reconstructed particle candidates,
as being due to the primary event or due to pileup.
No information is lost by subtracting information from the event, and instead
different weights are assigned according to a probability distribution.
The Pileup per Particle Identification algorithm~\cite{Bertolini:2014bba} is expected to lead to stable and unbiased
jet energy and estimation with optimal resolution and it is therefore one of the main candidates
to deploy in Run II to tame pileup.
Tightening further the cone used for jet clustering can also mitigate effectively the worsening of the resolution at the cost, 
however, of a poorer response and larger dependency on the modeling of hadronization and out-of-cone effects. 
More details can be found in~\cite{ATLAS-CONF-2014-018,CMS-PAS-JME-14-001}.

To maintain the performance of the identification of jets produced by heavy-flavored quarks (c and b),
the detectors will have to substitute their pixel detectors.
The additional pixel layer (IBL) installed last May in ATLAS 
is expected to lead to an improvement of the rejection of mis-tagged flavors by a factor of 2; 
the new CMS pixels will enter in operation in 2017 and are expected to improve the current performance, 
even when used in an average 50 pileup events scenario.

\subsection{Improvements from theory}
\label{subsec:theoryimprovements}

The preparation for the analysis of new data is also being done by updating
to the latest NLO with matched parton shower simulations for signals and backgrounds.
These are expected to yield more accurate predictions with reduced uncertainties from the choice
of the QCD scale. The inclusion of both non-resonant and resonant diagrams,
where the top quark width and the interference with the background continuum
are treated consistently, will improve our modeling of the signal.
New hadronizers, with improved decay tables, more recent underlying event tunes and
non-perturbative QCD models will also contribute for our understanding of the measurements to perform.
Ultimately, full NNLO predictions for differential distributions it both $\ttbar$ and single top events can be tested against larger data samples: charge asymmetry and top $\pT$ are two amongst the most relevant.
A complete overview has been given during this conference in~\cite{Seidel}.

\section{Crossing the precision frontier}
\label{sec:precisionfrontier}

\subsection{Production cross sections}
\label{subsec:productioncrosssections}

First day measurements will be focused on top quark production 
which, for $\ttbar$ production, 
is expected to increase by a factor of 5 (10) inclusively (for masses above 1 TeV).
It is important to notice that current uncertainties can only be partially improved from the experimental side as the main ones (luminosity and beam energy) will hardly improve with higher pileup and center of mass energy. From the theory side an improved understanding of the signal acceptance and single top modelling will benefit these measurements.
On the other hand, cross section ratios have the benefit of yielding partial cancellation of some of the uncertainties. Moreover, optimizing the measurement ot the ratio of cross sections 
at different energies (given the data available for the 8 and 7 TeV runs)
is expected to contribute to a better knowledge of the NNLO PDFs, in particular at large x,
given the fast rise in the $\ttbar$ cross section as function of the invariant mass of the di-parton system. Separating the gg and qq production, e.g. by associated W production tagging, will also be of interest.

The cross section for single top quark production is expected to increase less steeply than the $\ttbar$ one, due to PDFs. Thus, improving the current uncertainties will be challenging.
Exploring tighter phase space requirements and differential measurements
may help constraining more effectively the current uncertainties yielding
more precise measurements of $V_{\rm tb}$, $\mtop$, and other top properties, as well
as the PDF dependency of the cross section.
As for $\ttbar$, a global understanding of the top production as a whole
including resonant and non-resonant contributions will be mandatory in the new runs.
Establishing the associated production of t+Z will also be possible with new data 
and an important achievement to understand further the backgrounds for $\ttbar$+Z and $\ttbar$+H, amongst others.

In both cases, $\ttbar$ and single top, differential measurements, reported in fiducial regions,
provide crucial tests of pQCD, PDFs or alternative models.
Furthermore they have the potential of reducing the theory systematics and experimental biases if in-situ fits are performed. 
In this context establishing more precisely the $\pT$ of the top quark in $\ttbar$ events 
will be of particular importance as it impacts many precision measurements as well as BSM searches.
In general, if the underlying mechanism of production and decay of tops is established precisely enough,
other distributions related to charge asymmetry, spin correlations, missing transverse energy, amongst others, 
have the potential to constrain further BSM contributions.

\subsection{A global overview of the couplings}
\label{subsec:globaloverviewofcouplings}

Couplings of the top are naturally present at production and decay.
The tWb vertex structure has been found so far to be consistent with the SM 
but there is potential to improve its measurement
by combining different results obtained with $\ttbar$ and single top events.
With more data and a global overview fit we have to potential to measure all vectorial 
and dipole terms and also possible phases of the Lagrangian.

The necessary statistics needed to measure the 
couplings of the top quark to neutral vector bosons (Z, $\gamma$) will be achieved during Run II.
Given $\delta g_{\ttbar Z}\approx 10\%(1\TeV/\Lambda)^2$,
it is expected that at the end of the HL-LHC BSM contributions up to $\Lambda\sim\mathcal{O}(\TeV)$
are tested from the measurement of the coupling to the Z.
Comparable reach is expected for the case of the $\gamma$.

Similarly, BSM contributions to FCNC,
have the potential to be severely constrained with more data.
In these cases, discovery stories of FCNC with top using $\mathcal{B}(t\to cH)$,
with $H\to\gamma\gamma$,
or $\mathcal{B}(t\to qZ)$ in $\ttbar$ events
are expected to be possible in the HL-LHC scenario,
e.g. if $\mathcal{B}(t\to qZ)>0.02\%$.
In case of no discovery, FCNC searches will place stringent limits on BSM contributions
up to an energy scale significantly larger than the one the LHC will be able to produce directly.
The expected reach of the ATLAS detector for Run II and the HL-LHC
in limiting FCNC in the qZ and q$\gamma$ final states is summarized in
Fig.~\ref{fig:fcncfigure}.

The main interest of the next runs is, nevertheless, 
the measurement of the coupling to the Higgs boson. 
By combining different final states ($b\bar{b}$, $\gamma\gamma$, multi-leptons from W, Z or $\tau$) 
from $\ttbarH$ associated production,
it will be possible to measure the top Yukawa coupling with an uncertainty 
O(15\%) / O(10\%) after Run II / the HL-LHC. 
Interestingly, the current analyses observe an excess in the multi-lepton channel 
which will be important to explore with more data, as many BSM models predict these type of final states produced with top quarks.

A global overview of the top quark couplings, 
aggregating the information obtained from different final states and analyses,
as well as providing comparisons with meaningful benchmarks will be needed in the near future.
An example, using NLO computations, has been presented during this conference by~\cite{Zhang}.

\subsection{Mass and width}
\label{subsec:massandwidth}

\begin{figure}[h]
\begin{minipage}{0.48\textwidth}
\includegraphics[width=0.9\textwidth]{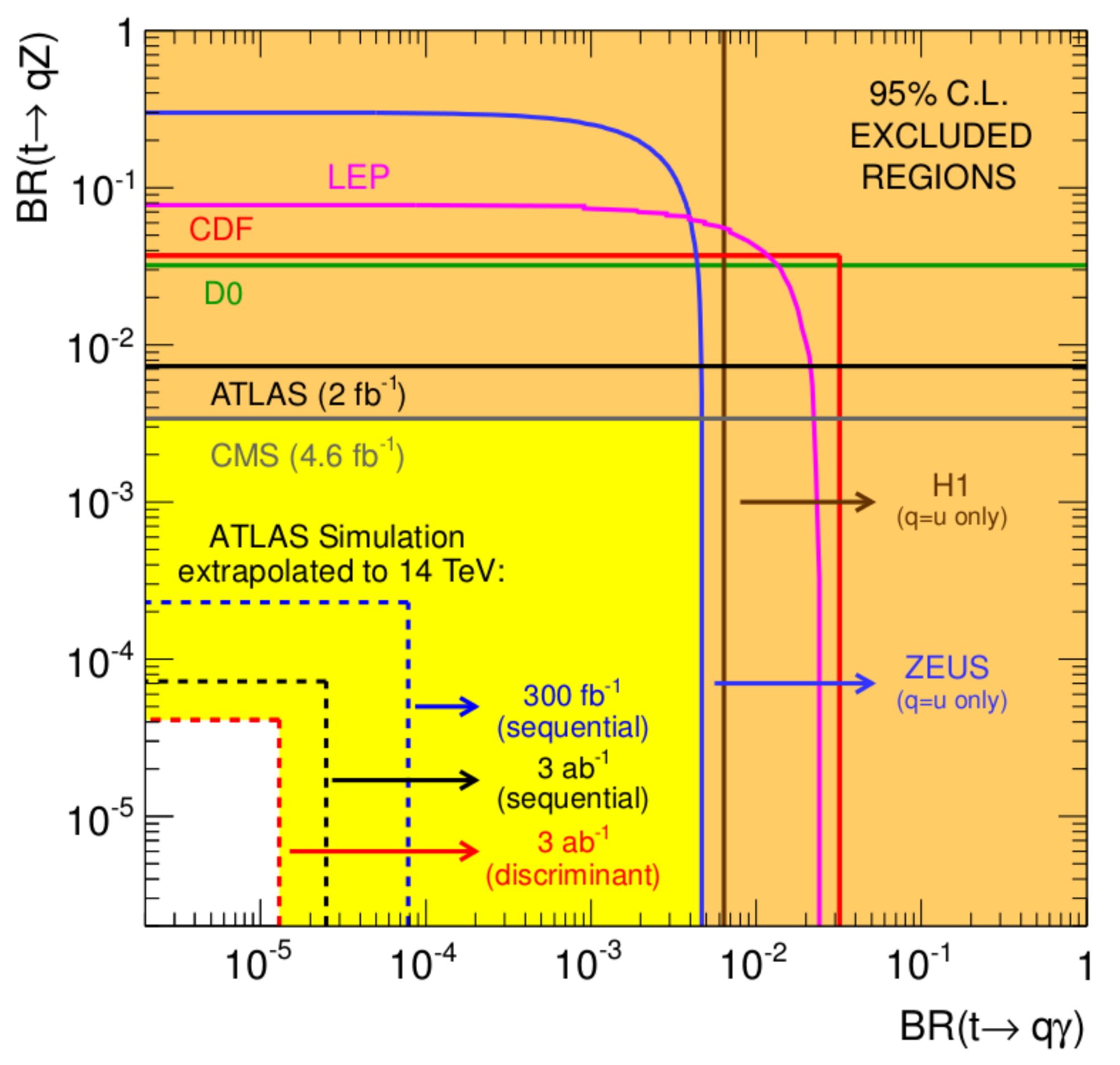}
\caption{\label{fig:fcncfigure}
Representation of the 95\% CL observed limits on the $\mathcal{B}(t\to\gamma q)$ vs. 
$\mathcal{B}(t \to Zq)$ plane are shown as full lines for the LEP, ZEUS, H1, D0, CDF, ATLAS and CMS collaborations. 
The expected sensitivity at ATLAS after $300\fbinv$ and $3\abinv$ are also represented by the dashed lines. 
From~\cite{ATLAS:2013hta}.}
\end{minipage}\hspace{0.02\textwidth}%
\begin{minipage}{0.48\textwidth}
\includegraphics[width=0.99\textwidth]{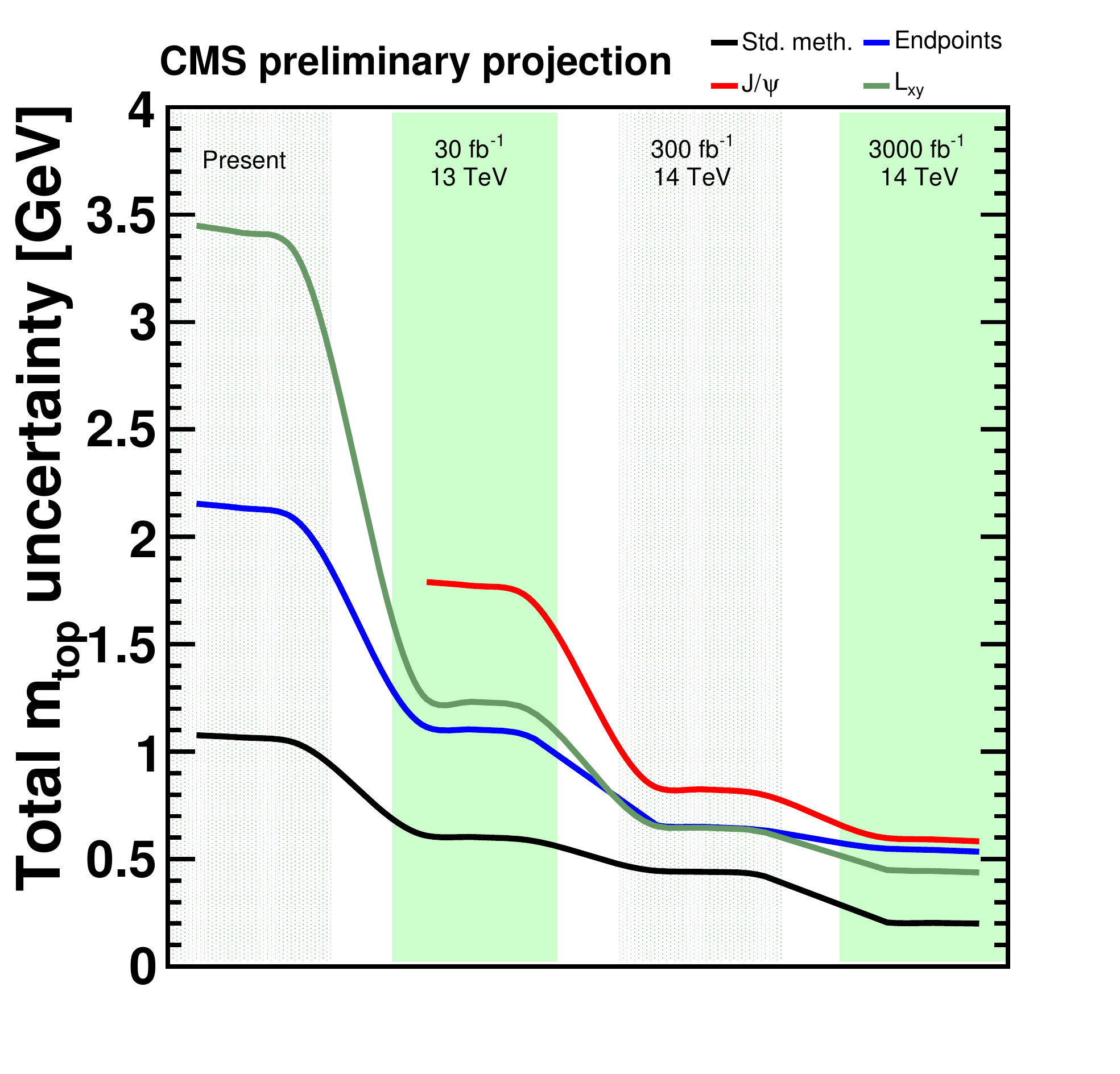}
\caption{\label{fig:mtopfigure} CMS projection of the top-quark-mass precision (in GeV) obtained with different methods, 
for various integrated luminosities. From~\cite{CMS-PAS-FTR-13-017}.
}
\end{minipage} 
\end{figure}

Current measurements of the mass attempt to make a step forward in both
{\it precision}, by expanding the number of variables which can be constrained in situ or 
disentangling the differential spectrum of the inclusively calibrated measurement; and
{\it illumination} of the nature of the measured parameter, by exploring alternative methods, different final states.
The standard measurements, initiated at the Fermilab Tevatron,
are expected to yield the most powerful results, from the experimental point of view.
Including in-situ constraints, higher order corrections from theory and improving the UE and fragmentation 
tunes with dedicated measurements in $\ttbar$ events,
are all expected to help reducing further the uncertainty
and contribute to clear up the road towards
a final uncertainty close to $\Lambda_{QCD}$ by the end of the HL-LHC.
Alternative measurements such as the B-hadron lifetime technique, spectra endpoints 
and based on the invariant mass of $\ell+J/\psi$ system are expected to attain sub-GeV uncertainties but, 
due to tighter requirements they are note expected to attain the same level of uncertainty.
Figure~\ref{fig:mtopfigure} summarizes the expected evolution of the uncertainty attained by different techniques.
A more detailed discussion can be found in~\cite{CMS-PAS-FTR-13-017}.

Intrinsically related to $\mtop$ is the width of the top quark.
A direct measurement from the mass shape, 
or the exploration of the production cross section where off-shell tops dominate 
may yield further insight on this fundamental property of the top quark.
Both techniques are expected to be feasible with larger datasets 
and also with a deeper understanding of the inclusive $\wwbbbar$ production from the theory side~\cite{Winter}.

\section{Unveiling the landscape of new physics}
\label{sec:unveilingthelandscapeofnewphysics}

Precision measurements and direct searches with top quarks strengthen each other 
helping bringing to light the BSM landscape. 
Although covering the different signatures which will be sought for 
with the next runs is out of scope of this manuscript it is important to refer
how top quark events can contribute to probe effectively the naturalness concept at the LHC.

Stop production through gluino-pair- or direct-production have the largest potential to
yield discovery stories in the next runs.
Discovery can occur both in cases where the mass splitting is large, 
with SUSY partners near the TeV scale,
as well as in cases where the $\ttbar$ plus missing energy signature would be more elusive, 
due to compressed SUSY spectra. 
The full hadronic and single lepton channels may drive the story of discovery due to its larger branching ratio
but other final states will help making a more solid case. 
The mass spectra of SUSY-like partners is expected to be covered up to 1.2 TeV (2 TeV) for stops (gluinos).
In this scenario the fine tuning of the theory would be tested up to $\mathcal{O}(1/100)$.
In a composite Higgs scenario, searching for extra symmetries, dimensions or exotic top partners
will test naturalness. Z' and Kaluza-Klein gluons will be probed up to a mass of 6 TeV,
while exotic top partners, such as vector-like-quarks will be explored in a mass range similar to the one explored for the stops.
However, BSM may be elusive and only in reach
by a systematic measurement of the differential spectra or by resigning from preferential coupling 
to b quarks if non-trivial flavor mixing occurs. 
Therefore no stone must be left unturned when looking directly for a sign of new physics
with top quarks.

\section{Final remarks}
\label{sec:conclusions}

Although both ATLAS and CMS collaborations are still finalizing the Run I data analyses, 
more data is clearly needed to take the next step 
towards understanding further the nature of the top quark and if it couples to BSM sectors.
Both experiments are working intensively to restart with similar, or improved performance, in harsher data-taking conditions.
Detector upgrades, improved reconstruction and calibration techniques, 
making use of the latest theory tools are some of the examples chosen in this run-through.
Precise cross section measurements, 
a global view of the couplings of the top, 
the ultimate top mass and testing naturalness 
are essential components of a long list which is expected to be achieved 
after scrutinizing the LHC Run II and the HL-LHC data.

\ack

The author wishes to acknowledge the help from R. Schwienhorst,
C. Tancredi, A. Lister, M. Owen, A. Meyer, M. Mulders and A. Giammanco 
for their help preparing the review here summarized.
The organizers of TOP2014 are also thanked for the successful workshop.

\end{document}